\documentclass[]{spie}  

\usepackage{amsmath,amsfonts,amssymb}
\usepackage{graphicx}
\usepackage[colorlinks=true, allcolors=blue]{hyperref}
\usepackage{caption}
\usepackage{subcaption}

\title{Simulating medium-spectral-resolution exoplanet characterization with SCALES angular/reference differential imaging}

\author[a]{Aditi Desai}
\author[a]{Stephanie E. Sallum}
\author[b]{Ravinder Banyal}
\author[c]{Natalie Batalha}
\author[d]{Natasha Batalha}
\author[e]{Geoff Blake}
\author[f]{Tim Brandt}
\author[d]{Zack Briesemeister}
\author[e]{Katherine de Kleer}
\author[g]{Imke de Pater}
\author[h]{Josh Eisner}
\author[i]{Wen-fai Fong}
\author[d]{Tom Greene}
\author[j]{Mitsuhiko Honda}
\author[c]{Isabel Kain}
\author[i]{Charlie Kilpatrick}
\author[a]{Mackenzie Lach}
\author[k]{Mike Liu}
\author[c]{Bruce Macintosh}
\author[a]{Raquel A. Martinez}
\author[e]{Dimitri Mawet}
\author[h]{Brittany Miles}
\author[l]{Caroline Morley}
\author[m]{Diana Powell}
\author[n]{Patrick Sheehan}
\author[c]{Andrew J. Skemer}
\author[o]{Justin Spilker}
\author[c]{R. Deno Stelter}
\author[p]{Jordan Stone}
\author[b]{Arun Surya}
\author[b]{Sivarani Thirupathi}
\author[h]{Kevin Wagner}
\author[q]{Yifan Zhou}
\affil[a]{UC Irvine, Irvine, CA, USA}
\affil[b]{Indian Institute of Astrophysics, Koramangala, Bengaluru, India}
\affil[c]{UC Santa Cruz, Santa Cruz, CA, USA}
\affil[d]{National Aeronautics and Space Administration, USA}
\affil[e]{California Institute of Technology, Pasadena, CA, USA}
\affil[f]{UC Santa Barbara, Santa Barbara, CA, USA}
\affil[g]{UC Berkeley, Berkeley, CA, USA}
\affil[h]{University of Arizona, Tucson, AZ, USA}
\affil[i]{Northwestern University, Evanston, IL, USA}
\affil[j]{Okayama University of Science, Okayama, Japan}
\affil[k]{University of Hawaii, Honolulu, HI, USA}
\affil[l]{UT Austin, Austin, TX, USA}
\affil[m]{University of Chicago, Chicago, IL, USA}
\affil[n]{National Radio Astronomy Observatory, Socorro, NM, USA}
\affil[o]{Texas A\&M University, College Station, TX, USA}
\affil[p]{US Naval Research Laboratory, Washington, D.C., USA}
\affil[q]{University of Virginia, Charlottesville, USA}

\begin{document} 
\maketitle

\begin{abstract}
SCALES (Slicer Combined with Array of Lenslets for Exoplanet Spectroscopy) is a 2 - 5 micron high-contrast lenslet-based integral field spectrograph (IFS) designed to characterize exoplanets and their atmospheres. The SCALES medium-spectral-resolution mode uses a lenslet subarray with a 0.34 x 0.36 arcsecond field of view which allows for exoplanet characterization at increased spectral resolution. We explore the sensitivity limitations of this mode by simulating planet detections in the presence of realistic noise sources. We use the SCALES simulator $\texttt{scalessim}$ to generate high-fidelity mock observations of planets that include speckle noise from their host stars, as well as other atmospheric and instrumental noise effects. We employ both angular and reference differential imaging as methods of disentangling speckle noise from the injected planet signals. These simulations allow us to assess the feasibility of speckle deconvolution for SCALES medium resolution data, and to test whether one approach outperforms another based on planet angular separations and contrasts.
\end{abstract}


\newpage

\section{INTRODUCTION}
\label{sec:intro}  

SCALES (Slicer Combined with Array of Lenslets for Exoplanet Spectroscopy) is a high-contrast integral field spectrograph (IFS) designed to characterize exoplanets and their atmospheres. It is a lenslet-based instrument that allows for observations to be made in the 2 - 5 micron range. Exoplanet characterization may be carried out at increased spectral resolution using the SCALES medium-spectral-resolution mode, which uses a lenslet subarray with a 0.34 x 0.36 arcsecond field of view in series with an image slicer.\cite{scales}

Exploring the sensitivity limitations of the SCALES medium-resolution mode is advantageous as it allows us to assess the feasibility of planet characterization using this mode, as well as to quantify of the effects of various systematic noise sources. We simulate planet detections in the presence of realistic noise sources using the SCALES simulator $\texttt{scalessim}$ to generate high-fidelity mock observations of planets while varying planet temperature, planet-star separations, and number of parallactic angles ($\eta$).\cite{scalessim} The images include speckle noise from their host stars, as well as other atmospheric and instrumental noise effects. We then employ angular and reference star differential imaging to recover the planet signals.

\section{METHODS}
\label{sec:methods}

    \begin{figure}
        \centering
        \begin{subfigure}[!b]{0.45\textwidth}
            \centering
            \includegraphics[height=5cm]{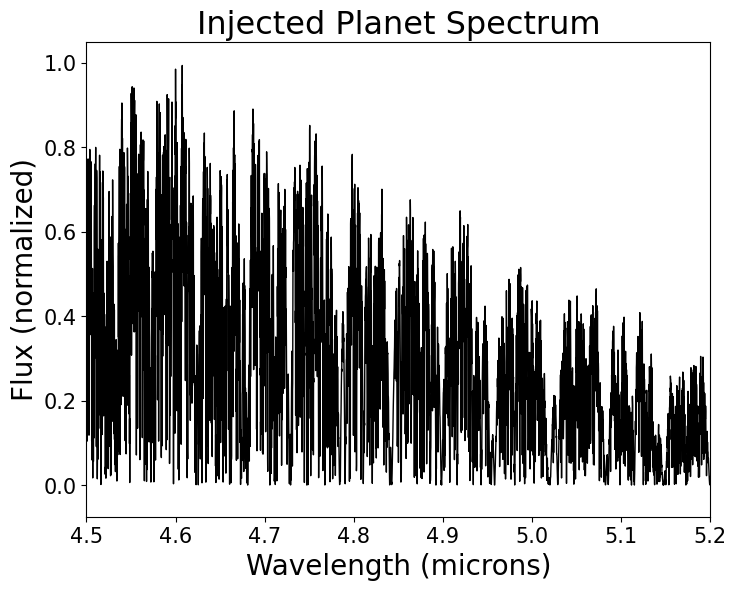}
            \caption{Injected spectrum of a 500 K planet.}
            \label{fig:inj500}
        \end{subfigure}
        \begin{subfigure}[!b]{0.45\textwidth}
            \centering
            \includegraphics[height=5cm]{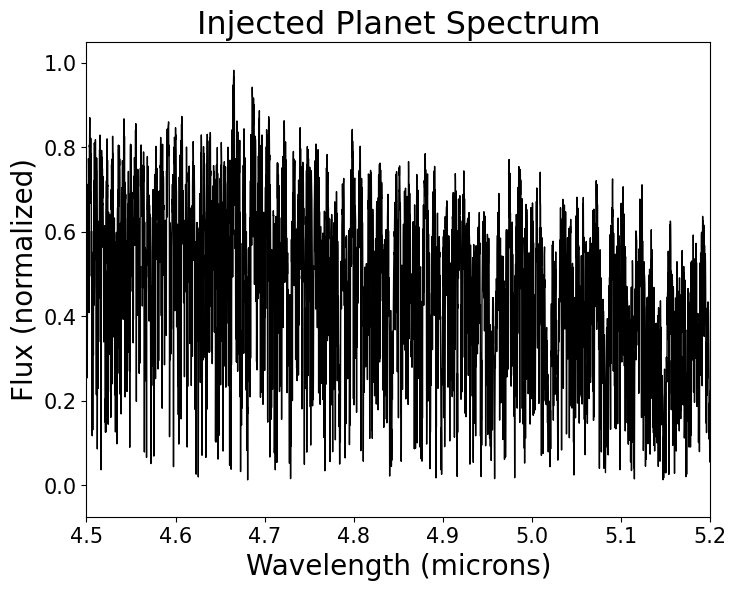}
            \caption{Injected spectrum of a 1000 K planet.}
            \label{fig:inj1000}
        \end{subfigure}
        \vspace{0.25cm}
        \caption{The injected Sonora Bobcat planet spectra we use in $\texttt{scalessim}$ when generating mock-observations.}
        \label{fig:inj}
    \end{figure}

In the SCALES medium-resolution mode, the field of view is not centered on the planet host star. Instead, the instrument centers on the exoplanet, following the star-planet system's parallactic rotation. This provides a series of images containing the planet as well as noise from the star that differs as its position with respect to the planet varies. We are able to simulate this with $\texttt{scalessim}$ by injecting a planet spectrum (e.g. Fig.~\ref{fig:inj}) and the spectrum of a host star, and cropping the scene to generate mock observations of that planet over a range of $\eta$, given a fixed planet-star separation. Once the images are generated, we can use angular and reference-star differential imaging to process the images and extract the planet signal.

All the simulations for this project were carried out for SCALES' medium-resolution M band mode, which covers the 4.5 micron to 5.2 micron range and has a resolving power of $R=7000$. We first inject a Sonora Bobcat planet spectrum for a 500 K or 1000 K planet.\cite{sb} Each of these planets is placed at a 500 mas or 800 mas separation from its 3000 K host star, which is modeled using a PHOENIX spectrum.\cite{ps} For each combination of planet temperature and planet-star separation, we generate a spectrally oversampled datacube of images at 34001 wavelengths and 13 or 21 parallactic angles over a range of $\eta=-60^{\circ}$ to $\eta=60^{\circ}$. We then rebin the oversampled cube according to $R = 7000 = \lambda / \Delta \lambda$ where $\lambda=\frac{5.2+4.5}{2}=4.85$ microns, and finally employ angular and reference-star differential imaging separately on each rebinned datacube.

\subsection{Angular Differential Imaging}
\label{sec:adi}

Angular differential imaging (ADI) is a high-contrast imaging technique that facilitates the detection of companions by reducing quasi-static speckle noise. In traditional ADI, observations of the host star are made with the instrument derotater turned off, allowing the the field of view to rotate with respect to the instrument. A reference point-spread-function (PSF)  that captures the average speckle noise in each image is constructed by selecting appropriate images from the same sequence. This reference PSF is subtracted from each image, and the residual images are rotated to align the field and then combined.\cite{adi}

\begin{figure}
        \centering
        \begin{subfigure}[ht]{0.45\textwidth}
            \centering
            \includegraphics[height=7.4cm]{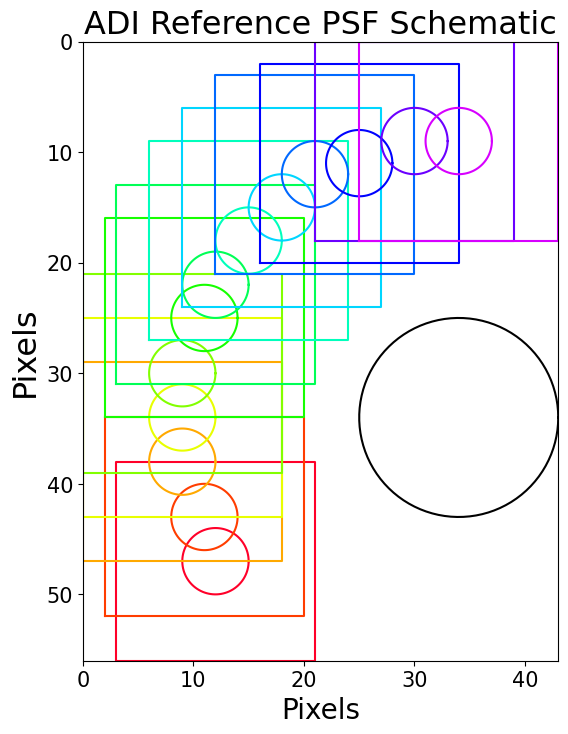}
            \caption{Schematic of ADI reference PSF construction. The squares represent the images, the circles within the squares represent the exclusion zone for each image in the sequence, and the black circle represents the host star PSF. The color progression indicates that each image is taken at a different value of $\eta$.}
            \label{fig:adischem}
        \end{subfigure}
        \hspace{0.2cm}
        \begin{subfigure}[ht]{0.45\textwidth}
            \centering
            \includegraphics[height=7.5cm]{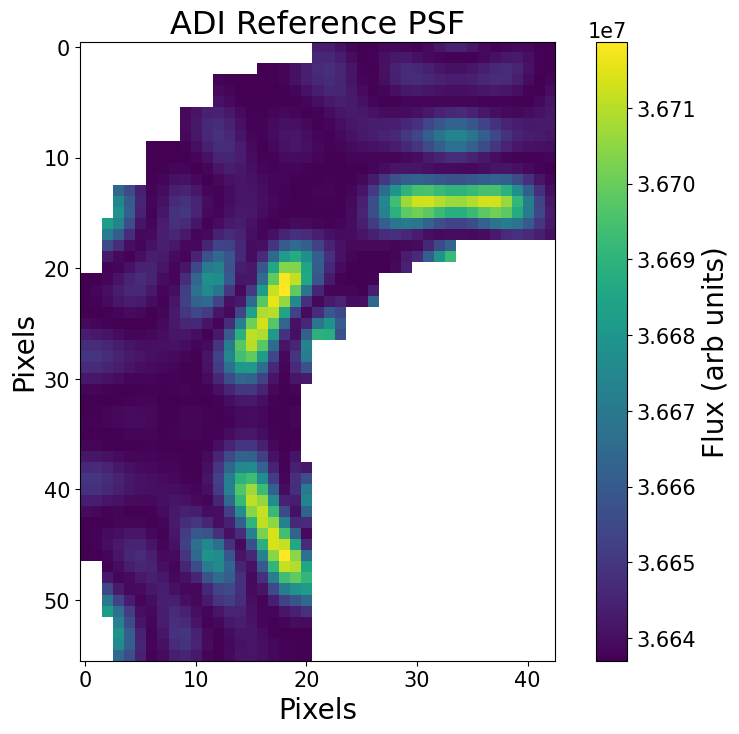}
            \caption{An example ADI reference PSF for a 1000 K planet around a 3000 K star. All light within the exclusion zone has been eliminated from each individual image in the datacube over all values of $\eta$ and all wavelengths, and the remaining noise was averaged over the range of $\eta$ values.}
            \label{fig:adipsf}
        \end{subfigure}
        \vspace{0.25cm}
        \caption{An ADI reference PSF constructed with 13 $\eta$ values and a 500-mas planet-star separation.}
        \label{fig:adi}
    \end{figure}

For images taken in SCALES medium-resolution mode, traditional ADI cannot be applied, as the field of view is not centered on the host star, so there is no rotation involved. To employ ADI for these images, we still generate a reference PSF, but we use a novel method of construction to do this. We first define an exclusion zone in each image that ignores light from the planet by blocking the light in that zone. For the purposes of this project, the exclusion zone is a fixed size across all wavelengths. We then calculate the mean of all overlapping images in a given region relative to the stellar PSF (e.g. one of the squares in Fig.~\ref{fig:adischem}, which gives us the desired reference PSF. We finally extract the planet by subtracting the reference PSF from the science image in that region, and then summing the residuals.
See Fig.~\ref{fig:adi} for a schematic of how the PSF is constructed and an example PSF.

\subsection{Reference-Star Differential Imaging}
\label{sec:rdi}

Reference star differential imaging (RDI) is another high-contrast imaging technique that uses images of other stars to build a model of the stellar PSF.\cite{rdi} For the purposes of this project, we use an idealistic scenario where the reference star used for processing is identical to the science target. The assumed observing scenario (i.e. airmass, precipitable water vapor, etc.) for the reference star is also assumed to be the same as that of the science target. For traditional RDI, principal component analysis (PCA) is often applied, but for this project, we achieved planet extraction only by subtracting the reference PSF from the science image at each parallactic angle.

    \begin{figure}
        \centering
        \begin{subfigure}[h]{0.24\textwidth}
            \centering
            \includegraphics[height=3cm]{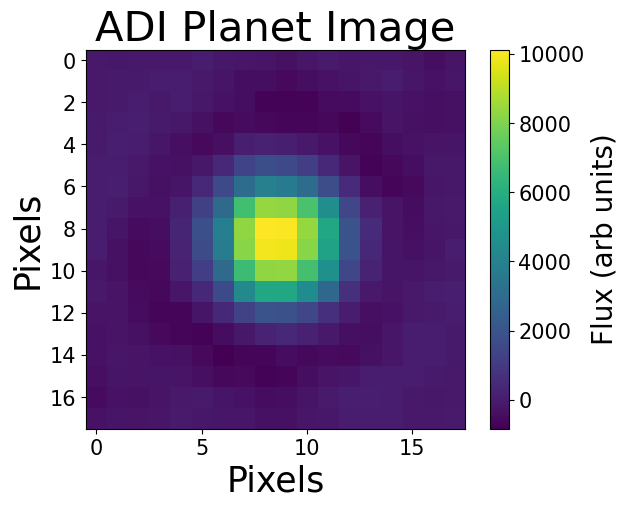}
            \caption{500 K, 500 mas, 13 $\eta$.}
            \label{fig:adi_500_13_500}
        \end{subfigure}
        \begin{subfigure}[h]{0.24\textwidth}
            \centering
            \includegraphics[height=3cm]{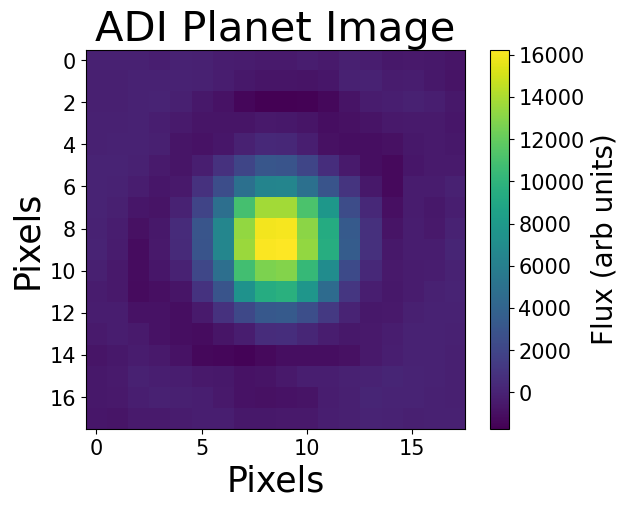}
            \caption{500 K, 800 mas, 21 $\eta$.}
            \label{fig:adi_500_21_800}
        \end{subfigure}
        \begin{subfigure}[h]{0.24\textwidth}
            \centering
            \includegraphics[height=3cm]{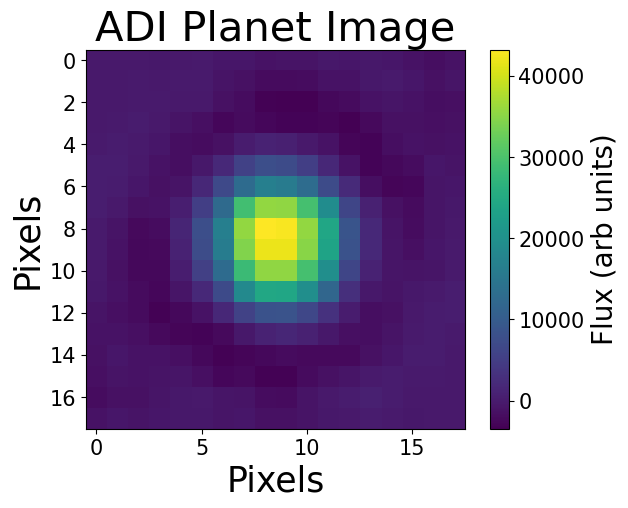}
            \caption{1000 K, 500 mas, 13 $\eta$.}
            \label{fig:adi_1000_13_500}
        \end{subfigure}
        \begin{subfigure}[h]{0.24\textwidth}
            \centering
            \includegraphics[height=3cm]{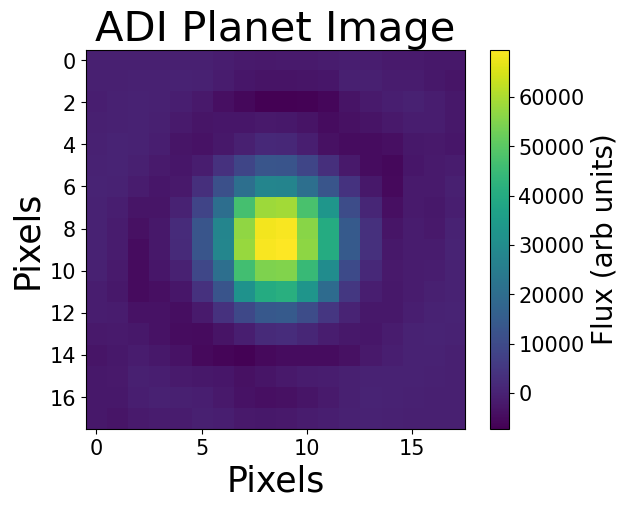}
            \caption{1000 K, 800 mas, 21 $\eta$.}
            \label{fig:adi_1000_21_800}
        \end{subfigure}
        \vspace{0.25cm}
        \caption{The images obtained after subtracting the ADI reference PSF from all science images and summing the residuals over all wavelengths for each combination of planet temperature, planet-star separation, and number of parallactic angles ($\eta$ values), which are specified in each panel.}
        \label{fig:adi_img}
    \end{figure}

    \begin{figure}
        \centering
        \begin{subfigure}[h]{0.24\textwidth}
            \centering
            \includegraphics[height=3cm]{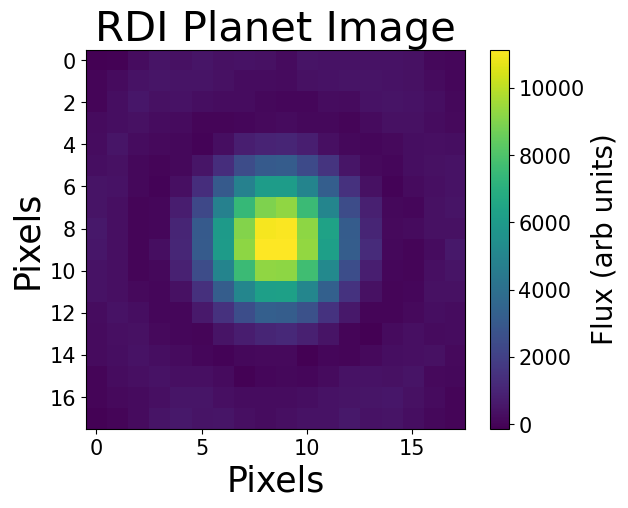}
            \caption{500 K, 500 mas, 13 $\eta$.}
            \label{fig:rdi_500_13_500}
        \end{subfigure}
        \begin{subfigure}[h]{0.24\textwidth}
            \centering
            \includegraphics[height=3cm]{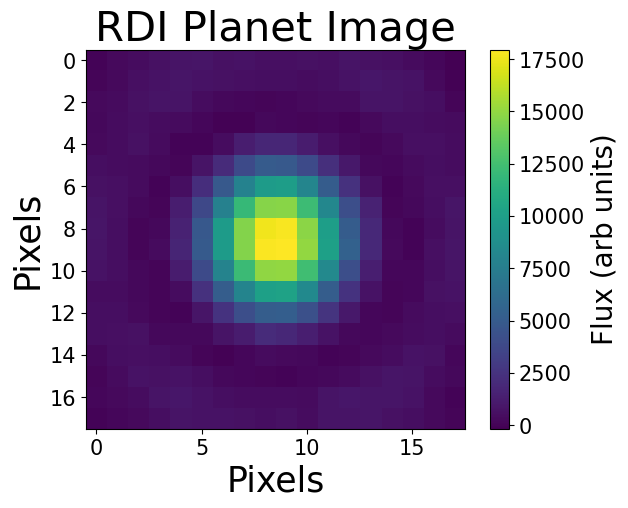}
            \caption{500 K, 800 mas, 21 $\eta$.}
            \label{fig:rdi_500_21_800}
        \end{subfigure}
        \begin{subfigure}[h]{0.24\textwidth}
            \centering
            \includegraphics[height=3cm]{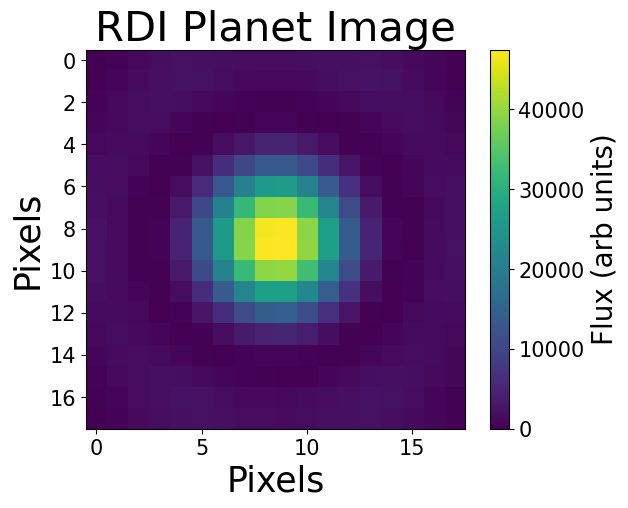}
            \caption{1000 K, 500 mas, 13 $\eta$.}
            \label{fig:rdi_1000_13_500}
        \end{subfigure}
        \begin{subfigure}[h]{0.24\textwidth}
            \centering
            \includegraphics[height=3cm]{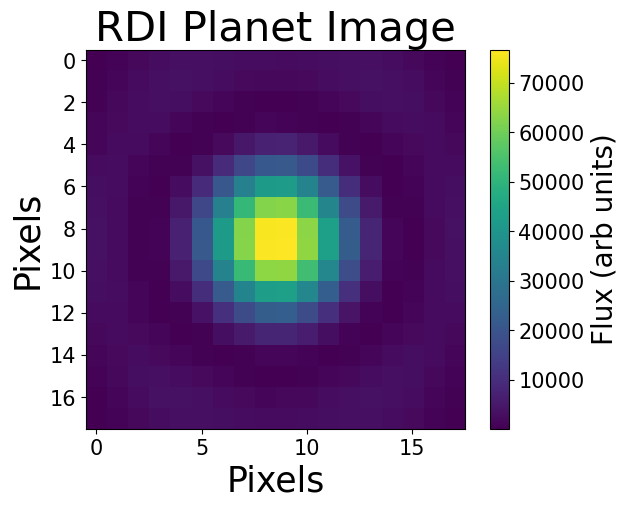}
            \caption{1000 K, 800 mas, 21 $\eta$.}
            \label{fig:rdi_1000_21_800}
        \end{subfigure}
        \vspace{0.25cm}
        \caption{The images obtained after subtracting the RDI reference PSF from all science images and summing the residuals over all wavelengths for each combination of planet temperature, planet-star separation, and number of parallactic angles ($\eta$ values), which are specified in each panel.}
        \label{fig:rdi_img}
    \end{figure}

    \begin{figure}
        \centering
        \begin{subfigure}[h]{0.95\textwidth}
            \centering
            \includegraphics[height=9cm]{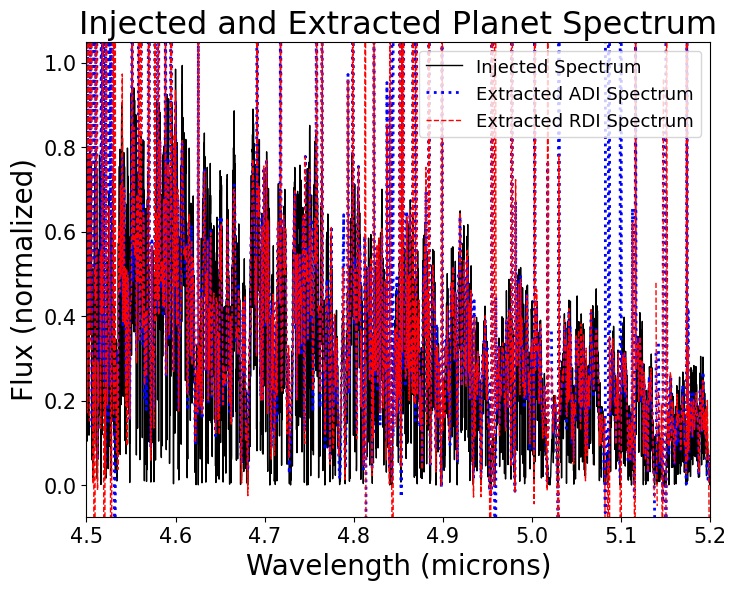}
            \caption{500 K, 500 mas, 13 $\eta$.}
            \label{fig:spec_500_13_500}
        \end{subfigure}
        \begin{subfigure}[h]{0.95\textwidth}
            \centering
            \includegraphics[height=9cm]{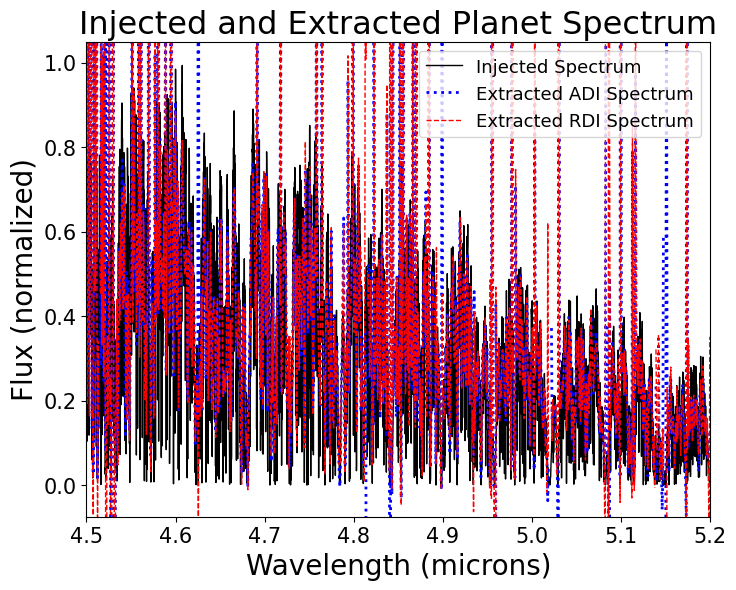}
            \caption{500 K, 800 mas, 21 $\eta$.}
            \label{fig:spec_500_21_800}
        \end{subfigure}
        \vspace{0.25cm}
        \caption{The spectra extracted after subtracting the ADI reference PSF (blue dotted line) or RDI reference PSF (red dashed line) from all science images and summing the residuals for a 500 K planet. Both extracted spectra are overlaid over the injected planet spectrum (black solid line), and the planet-star separation and number of parallactic angles ($\eta$ values) are specified in each panel.}
        \label{fig:500_spec}
    \end{figure}

    \begin{figure}
        \centering
        \begin{subfigure}[h]{0.95\textwidth}
            \centering
            \includegraphics[height=9cm]{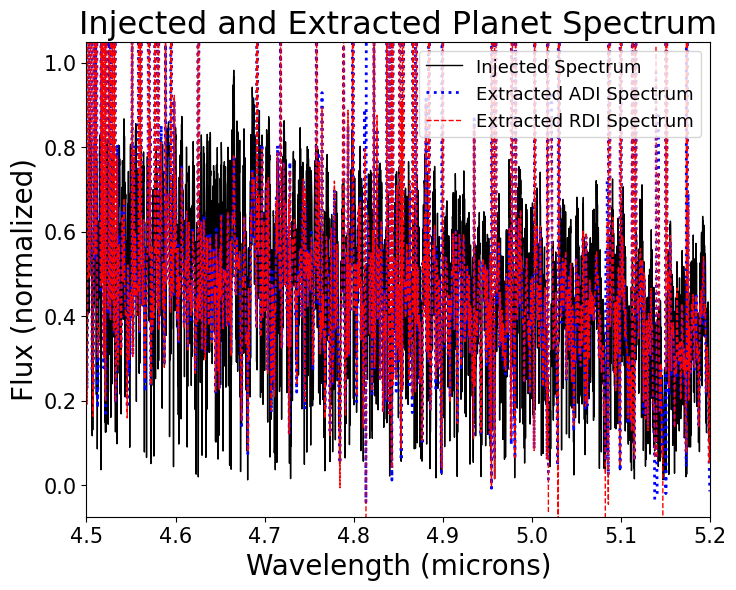}
            \caption{1000 K, 500 mas, 13 $\eta$.}
            \label{fig:spec_1000_13_500}
        \end{subfigure}
        \begin{subfigure}[h]{0.95\textwidth}
            \centering
            \includegraphics[height=9cm]{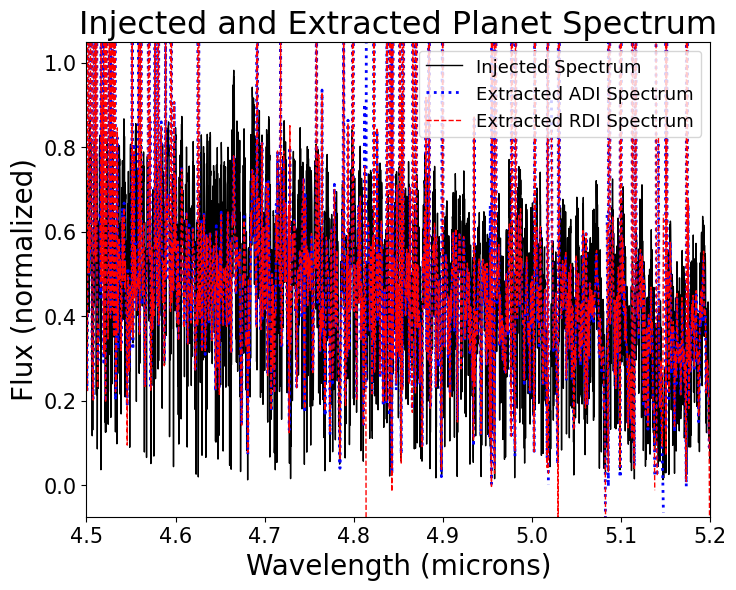}
            \caption{1000 K, 800 mas, 21 $\eta$.}
            \label{fig:spec_1000_21_800}
        \end{subfigure}
        \vspace{0.25cm}
        \caption{The spectra extracted after subtracting the ADI reference PSF (blue dotted line) or RDI reference PSF (red dashed line) from all science images and summing the residuals for a 1000 K planet. Both extracted spectra are overlaid over the injected planet spectrum (black solid line), and the planet-star separation and number of parallactic angles ($\eta$ values) are specified in each panel.}
        \label{fig:1000_spec}
    \end{figure}

\section{RESULTS}
\label{sec:results}


Figures~\ref{fig:adi_img}, ~\ref{fig:rdi_img}, ~\ref{fig:500_spec}, and ~\ref{fig:1000_spec} are the resulting images and spectra obtained after employing ADI and RDI on the mock observations generated by $\texttt{scalessim}$. All the images shown in Fig.~\ref{fig:adi_img} and Fig.~\ref{fig:rdi_img} have been summed over all wavelengths – however, detections can also be observed at individual wavelengths. We obtain the extracted spectra in Fig.~\ref{fig:500_spec} and Fig.~\ref{fig:1000_spec} by summing the entirety of every image at each wavelength and dividing out the sky transmission for an idealistic simulated telluric correction. Future work can be done to improve upon this by performing an aperture extraction or optimal extraction.\cite{opt}

In general, we find that both ADI and RDI appear to be effective for both large and small separations, as seen in Fig.~\ref{fig:adi_img} and Fig.~\ref{fig:rdi_img}. With both methods, the resulting images have comparable flux values for each combination of planet temperature and planet-star separation. Traditionally, ADI tends to be more effective for larger separations while RDI tends to be more effective for smaller separations.\cite{rdi}. In the case of RDI, the reason for the discrepancy between what is typically observed and the results of this work is likely that we have assumed an idealized scenario where the reference PSF and the science target PSF, as well as the observing conditions for both PSFs, are identical. In a more realistic scenario, we may obtain different extracted spectra.

Performing aperture extraction or optimal extraction should result in higher-quality spectra, but the simple approach carried out here (Fig.~\ref{fig:500_spec} and Fig.~\ref{fig:1000_spec}) still results in recovered spectra that agree with the injected signals for both separations across planet temperatures. Furthermore, the extracted spectra do not vary significantly regardless of whether ADI or RDI is employed. Several features, such as the overall decreasing trend in flux and the dips near 5.15 microns for both the 500 K and the 1000 K planets, can be matched with those in the injected spectra, regardless of which method is used. This similarity across methods indicates that the injected signal is indeed being recovered after ADI or RDI has been employed.

\section{DISCUSSION}

These simulations provide an initial assessment of how feasible it is to disentangle exoplanets from speckle noise in observations from the SCALES medium-resolution mode. The $\texttt{scalessim}$ mock observations demonstrate that both ADI and RDI are promising methods for processing and analyzing SCALES science images captured in this mode. We conclude that both methods may be successfully employed to recover the signal of an exoplanet. Further work must be done to determine the extent to which exoplanet characterization can be performed, and to compare the two techniques in a quantitative way. 

We must note that the work done here is idealized. In generating each scene that we use for both ADI and RDI, we assume a PSF that is not evolving in time, so no weighting is applied to take into account temporal changes in the PSF. We also assume that the reference star is identical to the exoplanet host star, and that the observing conditions and airmasses for both stars are the same for the RDI processing.

In the future, further refinement to these methods will address considerations that were not taken into account for this project. These include more variation in exoplanet temperature, variation in host star characteristics, temporal evolution of the PSF, the use of RDI reference stars that are not identical to the science target, and the use of more realistic PSF models. We may also couple these methods with spectral differential imaging (SDI) to leverage the wavelength information for improved stellar PSF deconvolution.\cite{sdi} Accounting for these additional criteria will allow for a more thorough comparison of these methods for processing SCALES medium-resolution mode data.

\acknowledgments 
We are grateful to the Heising-Simons Foundation, the Alfred P. Sloan Foundation, and the Mt. Cuba Astronomical Foundation for their generous support of our efforts. This project also benefited from work conducted under the NSF Graduate Research Fellowship Program. S.S. is supported by the National Science Foundation under MRI Grant No. 2216481. R.A.M is supported by the National Science Foundation MPS-Ascend Postdoctoral Research Fellowship under Grant No.~2213312.

\bibliography{report} 
\bibliographystyle{spiebib} 

\end{document}